# Features of phonon scattering by a spherical pore: molecular dynamics insight

Mykola Isaiev[1*], Nataliia Kyrychenko[1,2], Vasyl Kuryliuk[2], and David Lacroix[1]

[1]Université de Lorraine, CNRS, LEMTA, Nancy F-54000, France

[2]Faculty of Physics, Taras Shevchenko National University of Kyiv, 64/13, Volodymrska Str., Kyiv 01601, Ukraine

*Author to whom correspondence should be addressed: mykola.isaiev@univ-lorraine.fr

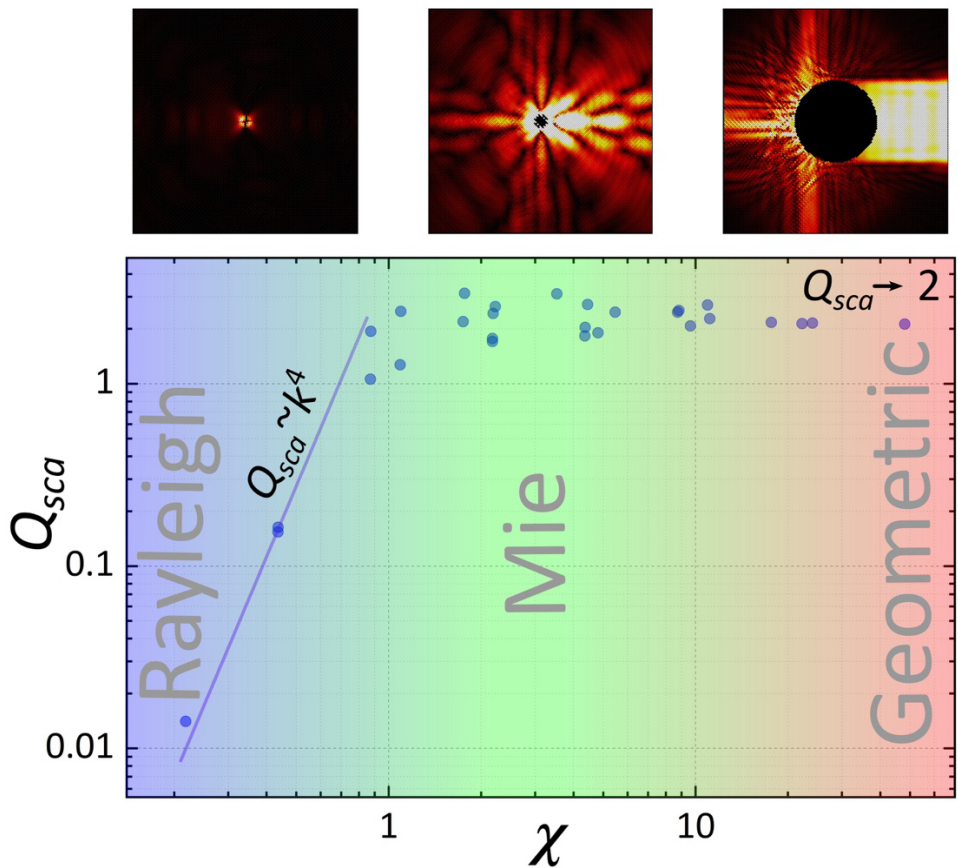

Abstract

There is still a gap in understanding phonons scattering by geometrical defects at the nanoscale, and it remains a significant challenge for heat transfer management in nanoscale devices and systems. In this study, we aim to explore the characteristics of phonon scattering by a single pore to gain insights into thermal transport in nanostructures. The paper outlines a methodology for assessing the spatial distribution of the magnitude of the radial, azimuthal, and polar components of the velocity of scattered phonons by a spherical pore. We demonstrated that the size parameter, commonly employed in electromagnetic wave scattering theory, is vital in determining the scattering regime. Specifically, we show that calculated scattering efficiency has the same pattern as one commonly obtained in classical wave scattering theory. However, we found that crystallographic directions are pivotal in shaping the scattering patterns, especially in the regions where scattering patterns are defined by the Mie resonances. This observation holds significance in understanding the influence of phonon coherence on thermal transport in nanostructured materials.

A comprehensive understanding of phonon dynamics at the nanoscale is essential for optimizing the performance, reliability, and efficiency of nanoscale devices and materials across a wide range of applications in nanoscience and nanotechnology. Recent micro- and nano-engineering progress allows us to manipulate materials at scales up to a few nanometers [1], where the surface-to-volume ratio defines the properties of the media. In such a way, the phonon scattering at the interface between different regions significantly reduces thermal transport properties and causes hot spots [2]. Introducing a high-density interfacial area challenges us to find more sophisticated approaches for cooling applications [3], [4]. Additionally, it should be noted that the interface is the source of the stresses arising in materials; such stress also significantly impacts the thermal transport properties of the media [5], [6], [7], [8].

A clear view of phonon dynamics is still a limiting bottleneck to establishing new pathways for the thermal engineering of various nanoscale devices and systems. Furthermore, understanding phonon scattering is crucial for developing new types of quantum computers based on phonon splitting [9].

Previously, the significant role of the phonons' wave properties on thermal transport at the nanoscale has already been stated. Indeed, such an effect became necessary when the size of the scattering object is less than the phonon wavelength [10], [11]. In this case, the explanation of the observed data requires the application of the approaches based on the Rayleigh and Mie scattering regimes [12], [13], [14], which are specific to the wave nature of the phonons. For instance, Rayleigh-based Ziman's scattering model [15] has proven to be sufficient for describing the experimental data [16] and correlate well with other simulations [17], [18] when the defect size is much less than the phonons' wavelength.

On the other hand, when the phonons' wavelength is much smaller than the characteristic defect size, the model based on geometric scattering can be applied [19]. This allows the development of a representation of the phonon mean free path for different inclusions in geometries and configurations [20], [21], [22]. Furthermore, phonons' behavior can be considered particle-like, and methods based on their dynamics, such as the Monte Carlo method for Boltzmann transport equation resolution, can be applied [23], [24], [25].

Several models were proposed to describe the cross-over from Rayleigh to geometric scattering [22], [23], [24]. However, some questions should still be addressed, specifically concerning the application of such approximations to crystalline anisotropic media and the possible impact of Mie resonance on the scattering cross-section.

Current progress in simulations allows us to look deeper inside the phonon's dynamics at the nanoscale. It is crucial to mention that multiscale simulations can decompose the impact of different aspects on integral thermal transport. Specifically, in this paper, we investigate phonon scattering by a pore to understand features of thermal transport in the media with nanoscale objects. For our simulations, as a modeled system, we consider phonon scattering in a system with a pore with different pore radii and wavelengths of phonons. Information regarding phonon scattering is essential for predicting thermal conductivity in highly porous media [7], [22], [26].

For such investigations, we implemented an approach based on wave-packet formalism combined with molecular dynamics [27]. This formalism recommended itself as a valuable tool for investigating phonon propagation in nanostructured materials [28], [29]. This approach was successfully applied to investigate the phonon scattering at the planar surfaces with different structuration [30], [31], [32]. These investigations are crucial for the revision of Ziman' model for the reflection behavior at random rough surfaces, which is crucial for the description of the phonon scattering at the planar interfaces.

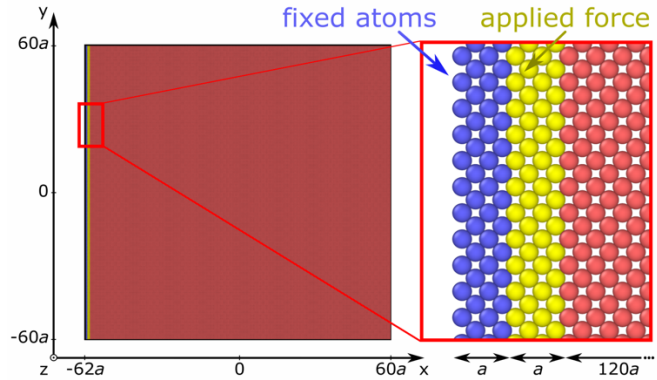

**Fig. 1.** Geometry of the reference system, which was created as monocrystalline diamond-like silicon with a lattice parameter equal to $a$ = 5.43 Å; the simulation domain for the reference system consists of 122×120×120 repetitions of the unit cell in x, y, and z directions, respectively. The coordinate system was chosen so that -62$a$ < $x$ < 60$a$ was centered to zero in the y and z directions.

For the simulations, we considered monocrystalline diamond-like silicon with a lattice parameter equal to $a$ = 5.43 Å; the simulation domain for the reference system consists of 122×120×120 repetitions of the unit cell in x, y, and z directions, respectively. The coordinate system was chosen so that -62$a$ < $x$ < 60$a$, and it was centered to zero in the y and z directions. A pore was created by cutting the atom inside a sphere with the radius $R$ = 1$a$, 2$a$, 5$a$, and 10$a$ centered in the point (0, 0, 0). The interactions between silicon atoms were calculated using Tersoff potential [33]. This potential is chosen to represent the anharmonic interactions between silicon atoms. Another potential, based on recently developed machine learning approach, can be also used [34]. The energy minimization procedure relies on the conjugate gradient algorithm to relax the system. The initial temperature of the system was set to 0 K to decompose the impact of phonon-phonon scattering and the scattering at the pore's edge.

For the creation of the phonon wave-packet (see **Fig.1** for the geometry of the initial system), the first layer of the unit cell in x direction ($-62a < x < -61a$), and for the second layer ($-61a < x < -60a$), the periodic force in x direction was applied:

$$F = F_0 \exp(2\pi i v t), \qquad (1)$$

where $F_0$ is the force magnitude, and $v$ is the frequency of modulation. Our investigation considered the following frequencies: 0.5, 1, 2, 2.5, 4, 5, and 10 THz. Such frequencies were chosen to match the proportionality between the quarter of the oscillation period and the timestep (0.5 fs). Such simulations were carried out for the reference system and all systems with an increased pore radius R. It should be noted that the anharmonic interaction is presented through the interactional potential despite the excitation force being chosen harmonic-like.

Such periodic force created a wave of elastic displacement and velocities with the wavelength $\lambda(v)$. The spatial distribution of the velocity field inside the reference system and the dependence of the wavelength on the frequency are presented in **SM2**. For analysis, we use the size parameter defined as follows:

$$\chi = \frac{2\pi R}{\lambda}. \qquad (2)$$

For the calculation of the velocity component in a wave scattered at the pore, we decompose the velocity field in the systems with the pore as follows:

$$\vec{V}_{pore} = \vec{V}_{incident} + \vec{V}_{scattered}, \qquad (3)$$

where $\vec{V}_{pore}$ is the total velocity field in the system with a pore, $\vec{V}_{incident}$ is the velocity field of the incident wave, and $\vec{V}_{scattered}$ is the velocity field of a scattered wave.

We assumed that the velocity field in the reference system ($\vec{V}_{reference}$) for corresponding frequency is equal to the velocity field in the incident wave due to the absence of scattering objects. It allows us to recalculate the scattering as follows:

$$\vec{V}_{scattered} = \vec{V}_{pore} - \vec{V}_{reference}. \qquad (4)$$

For our analysis in this letter, we decided to consider velocities in the radial, azimuthal and polar directions to the pore surface, which were recalculated as follows:

$$\vec{V}_{scattered} = \vec{V}_r \vec{e}_r + \vec{V}_\phi \vec{e}_\phi + \vec{V}_\theta \vec{e}_\theta, \tag{5}$$

where $\vec{V}_i = (\vec{V}_{scattered}, \vec{e}_i)$, $i = r, \phi, \theta$, and $\vec{e}_r, \vec{e}_\phi$, and $\vec{e}_\theta$ are the unit vectors in radial, polar, and azimuthal directions.

Finally, the amplitude of the radial component of the velocity field in scattered waves was analyzed. To simplify further consideration, we normalized the obtained amplitude by the magnitude of the incident wave:

$$v_i = \frac{\sqrt{V_i V_i^*}}{A}, \tag{6}$$

where $A$ is the magnitude of the excitation filed for the respective frequency.

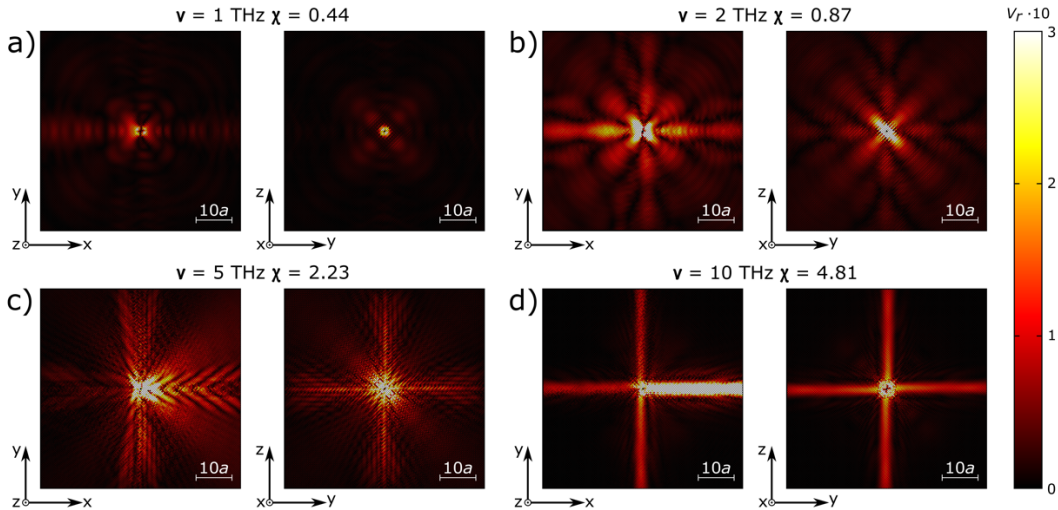

**Fig. 2.** The (x, y) and (y, z) slices of the spatial distribution of the radial component of the normalized velocity scattered at a pore with radius equal to 1$a$ for different modulation frequencies of the force: a) ν = 1 THz; b) ν = 2 THz; c) ν = 5 THz; d) ν = 10 THz. k-vector of the incident wave is direct along the x-axis

It should be noted that a similar procedure was also performed for the displacement. However, due to the specificity of the oscillation excitation, the spatial distribution of displacement fluctuates over some constant not equal to zero (see SM1), which complicates the harmonic analysis. Since both momentum and displacement may be considered equivalent to describing phonon properties, we further decided to consider only velocities. Nevertheless, the pattern for displacement and velocities are almost equivalent (see SM2).

Fig. 2 and 3 present the (y, z) and (x, y) slices of the selected molecular dynamic snapshots depicting atoms' positions for pore radii of 1$a$ and 2$a$, respectively. The slices (y, z) have dimensions of 54$a$ in y and z direction (-27$a$ < y, z < 27$a$), and a thickness is 2$a$ (-$a$ < x < $a$). The slices (x, y) have dimensions of 54$a$ in x and y directions (-27$a$ < x, y < 27$a$), and a thickness is 2$a$ (-$a$ < z < $a$). Each atom is color-coded based on the calculated velocity magnitude $v_r$.

SM2 presents the figures corresponding to other pore radii and the velocities component. Additionally, SM2 presents the corresponding snapshots for the displacement for the pore radius equal to 1$a$. The figures include information about the size parameter $\chi$ of all cases, which, according to continuous field theory, is vital in predicting the scattering regime [35].

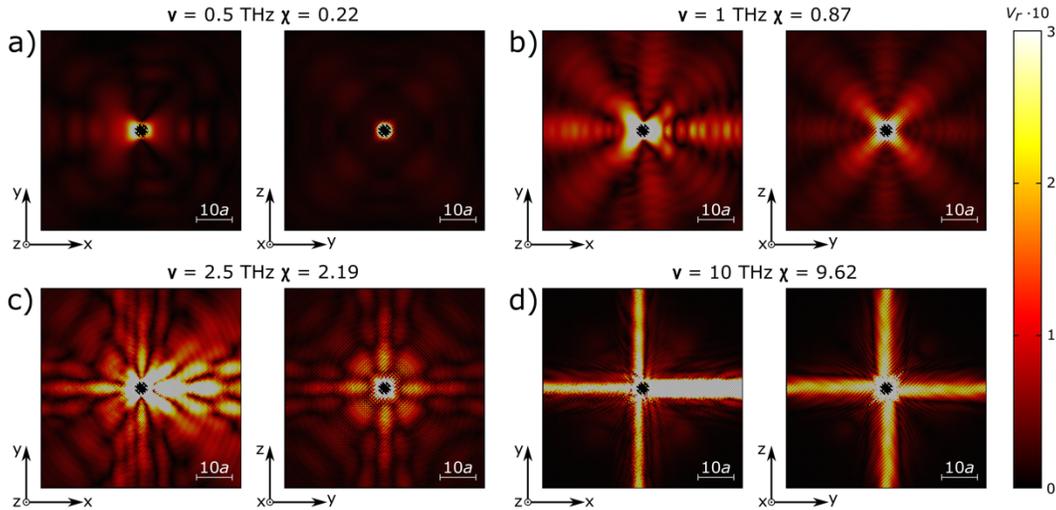

**Fig. 3.** The (x, y) and (y, z) slices of the spatial distribution of the radial component of the normalized velocity scattered at a pore with radius equal to 1$a$ for different modulation frequencies of the force: a) $v$ = 0.5 THz; b) $v$ = 1 THz; c) $v$ = 2.5 THz; d) $v$ = 10 THz. k-vector of the incident wave is direct along the x-axis

We observed an increased decay in the scattering field as the $\chi$ parameter decreases, aligning well with previously reported findings [7]. In the latter study, the authors emphasized the significance of necks between scattering features in describing the thermal transport properties of nanostructured media.

As the Fig.2-3 show, we observe different scattering regimes, starting from Rayleigh (for small values $\chi < 1$), Mie-like, and finally, geometric scattering (for larger values of $\chi$). The scattering diagrams calculated in the frame of Mie approximation for isotropic media are also presented in **SM3**. It is clear that the transition between different scattering regimes is smooth, and all respective regions will be discussed below in detail. It should be noted that there is a correlation between the results of Mie modeling and the results of MD simulations. Nevertheless, a substantial difference is observed due to the anisotropic media and near-field MD observation.

Figures 2-3 elucidate how the scattering diagram undergoes further modifications, influenced by phonon lifetimes in various directions. Notably, along directions (1, 0, 0) (parallel to the x, y, and z axes) and (1, 1, 0) (parallel to diagonals), the velocity field experiences less pronounced decay with distance from the pore compared to other directions due to the much higher lifetime of phonons propagating in these directions.

As one can see, for the small values of the size parameter, when the scattering is defined by Rayleigh and Mie regimes with wide angular dependence of the scattering field, the direction (1, 1, 0) is more pronounced. Furthermore, the wave behavior of the scattering fields is intriguing. A beam of the scattered field tends to expand following the Huygens–Fresnel principle, the resulting front is non-spherical due to the media's anisotropy. This phenomenon becomes apparent when the beam is narrow; as seen in Fig. 2c, distinctive fishbone-like patterns emerge.

It should be note that with increasing of $\chi$ we observe the transition from the Mie to the geometric scattering regime, and the respective patterns of the scattering field define by the reflection and shadowing of the scattering field. This leads to vanishing of the scattering in (1, 1, 0) direction.

It is crucial to acknowledge the occurrence of both the $v_\phi$ and $v_\theta$ components, which propagate radially, indicating a conversion from a purely longitudinal excitation field to a combination of longitudinal and transverse scattered fields. Our simulations revealed approximately equal energy distribution among these three components. Nonetheless, this is just a rough estimation, and further investigation is needed to examine this observation precisely.

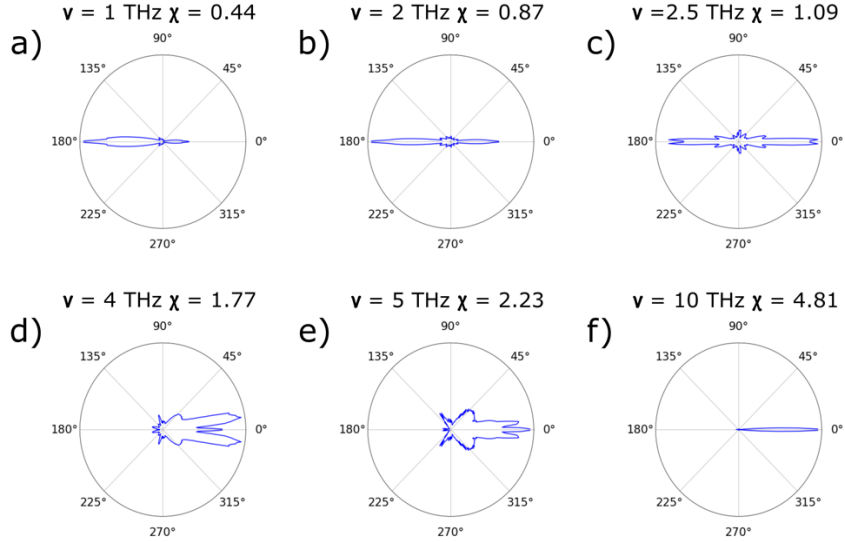

[36]
**Fig. 4.** Phase function ($f(\theta)$) calculated of the field scattered from the pore's edge with a radius equals to $1a$

The insights gained from the spatial distribution of the scattered field, as depicted in Figures 2-3 and SM2, can be readily extrapolated to other materials using the approach outlined in the paper. This paves the way for developing efficient systems to manage phonon transport. These systems could include technologies for scattering phonons with specific wavelengths, phonon membranes, and other innovative solutions geared towards precisely controlling and manipulating phonon scattering behavior.

For further analysis, we calculated, similar to classical scattering theory, the scattering diagrams with phase function averaged over the polar angle with the following equation:

$$f(\theta) = \frac{1}{2} 2\pi \langle (v_r^2 + v_\phi^2 + v_\theta^2) r^2 \rangle_{\phi, r=15a..25a}, \qquad (7)$$

where $\langle ... \rangle_{\phi, r=15a..25a}$ is the average over $\phi$ in frames from 0 to $2\pi$ and $r$ from $15a$ to $25a$. The factor 1/2 arises from the fact that in Eq. (6), the amplitude of the excited field is used instead of the value averaged over a period. In SM2, the scattering diagram for the specified case is calculated using this approach, and Mie theory is presented for comparison. Mie diagrams were calculated using the "miepython" package [36].

Observing the figure reveals a distinctive pattern: for low values of χ, the scattering field adopts a Rayleigh-like shape, featuring comparable back and forward scattering components. As $\chi$ increases, the leaf structures mimic Mie scattering, influenced by crystallographic orientations, which reduces the proportion of backward scattering. Ultimately, a predominant forward scattering emerges, characteristic of the geometric scattering regime.

For the generalization of the data computed for all frequencies and pores' radii, the scattering efficiency [37] function was calculated as follows:

$$Q_{sca} = \frac{1}{\pi R^2} \int_0^\pi f(\theta) \sin(\theta) d\theta. \tag{7}$$

Additionally, forward scattering efficiency was calculated as follows:

$$Q_{forward} = \frac{1}{\pi R^2} \int_0^{\frac{\pi}{2}} f(\theta) \sin(\theta) d\theta. \tag{8}$$

The resulting dependence of the total scattering efficiency and forward scattering efficiency are presented in Fig. 5 with respect to the size parameter. This dependencies exhibit a familiar pattern akin to those typically encountered within the framework of the Mie approach [37]. As one can see from the Fig.5, we can decompose three scattering regimes. For the values of the $\chi < 1$, we observe that the scattering field increase following $\sim k^4$ trend predicted by the Ziman scattering theory. Conversely, for χ > 10, the scattering efficiency tends towards a constant value close to 2, as predicted by classical wave scattering theory [37], corresponding to geometric scattering regime. In the transitional phase between Rayleigh and geometric scattering, fluctuations manifest due to the interplay between Mie resonances and the discrete direction of phonon propagation. Additionally, the comparison between total and forward scattering efficiency validates our observation regarding the increase in the proportion of forward scattering with the size parameter.

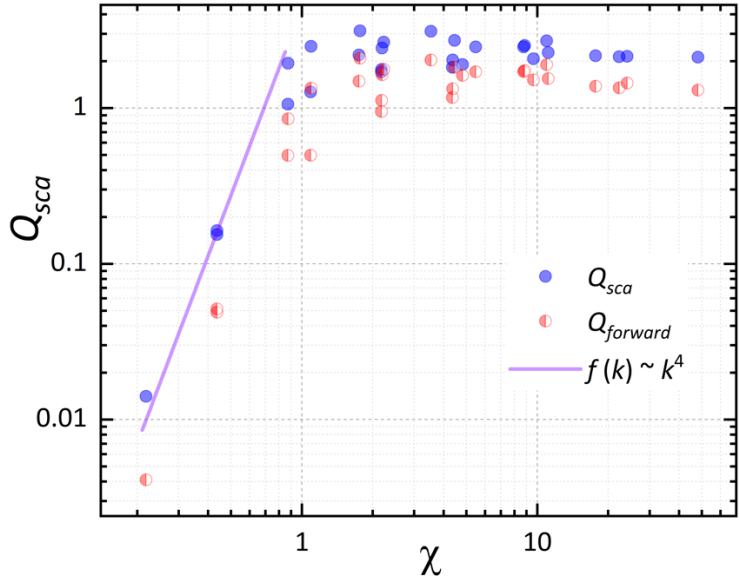

**Fig. 5.** Dependence of the total scattering efficiency and forward scattering efficiency with the size parameter

The results presented in this paper outline a method for evaluating the spatial distribution of the magnitude of the radial, azimuthal and polar components of the velocity for scattered phonons at the edge of a spherical pore. Based on the calculations of the spatial distribution of velocity field, we calculated scattering efficiency as a function of the size parameter ($\chi = 2\pi R/\lambda$), commonly used in scattering theory. We demonstrate, that the size parameter plays a determining role in defining the scattering regime. However, crystallographic directions are equally essential in shaping the scattering diagrams. This observation holds significance for understanding the impact of phonon coherence on thermal transport in nanostructured materials.

The methodology outlined for calculating scattering characteristics holds the promise of predicting material structures, such as pore/inclusion morphology, for managing phonon transport, including guiding or scattering phonons with specific wavelengths. Consequently, this study lays the groundwork for potential advancements in the thermal engineering of nanostructured materials. Such insights could prove instrumental in developing strategies for designing nanoscale devices with enhanced efficiency.

**Supplementary materials**

S1 shows the molecular dynamics snapshots of the positions of the particles, which are colored for the x-component of per-atom velocities. An average velocity profile of the x-component of velocity and its fitting by the sinus exists. Also, it proposes the example of the x-component of the displacement spatial distribution. The table in the S1 shows the obtained phonons' wavelength and size parameter for each frequency and pore size. S2 presents the (y, z) and (x, y) slices of the spatial distribution of the radial component of the normalized velocity scattered at a pore with radii equal to 1$a$, 2$a$, 5$a$, and 10$a$ and for different modulation frequencies of the force (1 THz, 2 THz, 2.5 THz, 4 THz, 5 THz and 10 THz). Also, the scattering diagram calculated with Eq. (7) and "miepython" [36] for different values of size parameter $\chi$ are included.

**Acknowledgment**

This paper contains the results obtained in the frames of the ANR project "PROMENADE," No. ANR-23-CE50-0008. Calculations were performed using HPC resources from GENCI-TGCC and GENCI-IDRIS (No. A0150913052); HPC resources were partially provided by the EXPLOR Center hosted by the Université de Lorraine. NK thanks the ERASMUS+ program between Taras Shevchenko National University of Kyiv and the Université de Lorraine for providing financial support for the internship.

**Data availability**

The data supporting this study's findings are available from the corresponding author upon reasonable request.

**Conflict of interest**

The authors declare the absence of a conflict of interest.

# Supplementary Materials
# Features of phonon scattering by a spherical pore: molecular dynamics insight

Mykola Isaiev[1*], Nataliia Kyrychenko[1,2], Vasyl Kuryliuk[2], and David Lacroix[1]

[1]Université de Lorraine, CNRS, LEMTA, Nancy F-54000, France

[2]Faculty of Physics, Taras Shevchenko National University of Kyiv, 64/13, Volodymrska Str., Kyiv 01601, Ukraine

*Author to whom correspondence should be addressed: mykola.isaiev@univ-lorraine.fr

## Supplementary materials 1

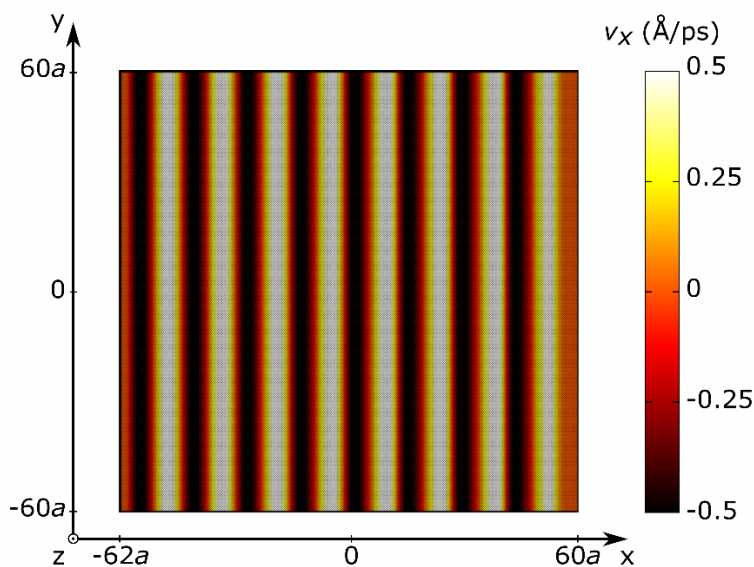

**Fig. S1.1.** The molecular dynamics snapshots showing the distribution of $v_x$ in the system at the end of the simulations for the modulation frequency equal to 1 THz

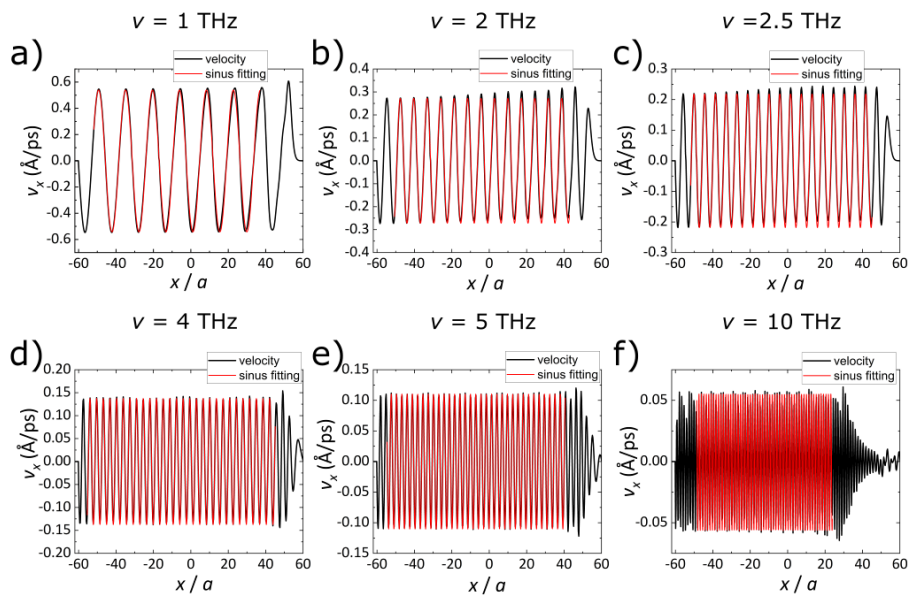

**Fig. S1.2.** Velocity distribution for different modulation frequencies of the force: a) v = 1 THz; b) v = 2 THz; c) v = 2.5 THz; d) v = 4 THz; e) v = 5 THz f) v = 10 THz; d) $R = 10a$. Black line molecular dynamics data, red line – sinus fitting

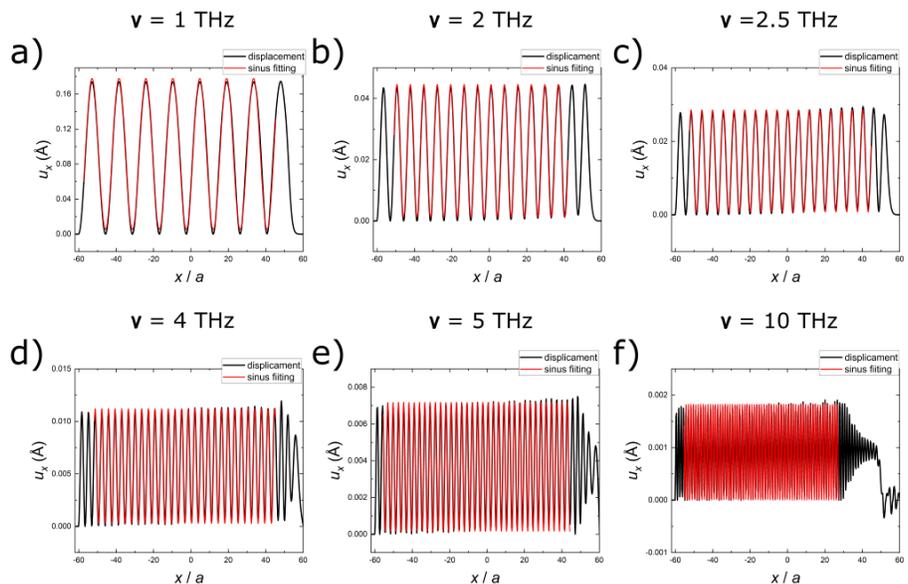

**Fig. S1.3.** Displacement distribution for different modulation frequencies of the force: a) v = 1 THz; b) v = 2 THz; c) v = 2.5 THz; d) v = 4 THz; e) v = 5 THz f) v = 10 THz; d) $R = 10a$. Black line molecular dynamics data, red line – sinus fitting

**Table S1.1.** Wavelegths of the excited waves under considered freaquancies and the size parameter ($\chi = \frac{2\pi R}{\lambda}$) for different pore radii

| ν (THz) | λ (Å) | $\chi$ | | | |
|---|---|---|---|---|---|
| | | R = 1a | R = 2a | R = 5a | R = 10a |
| 0.5 | 156.615 | 0.22 | 0.44 | 1.09 | 2.18 |
| 1 | 78.26 | 0.44 | 0.87 | 2.18 | 4.36 |
| 2 | 39.03 | 0.87 | 1.75 | 4.37 | 8.74 |
| 2.5 | 31.17 | 1.09 | 2.19 | 5.47 | 10.95 |
| 4 | 19.31 | 1.77 | 3.53 | 8.83 | 17.66 |
| 5 | 15.33 | 2.23 | 4.45 | 11.13 | 22.26 |
| 10 | 7.09 | 4.81 | 9.62 | 24.06 | 48.12 |

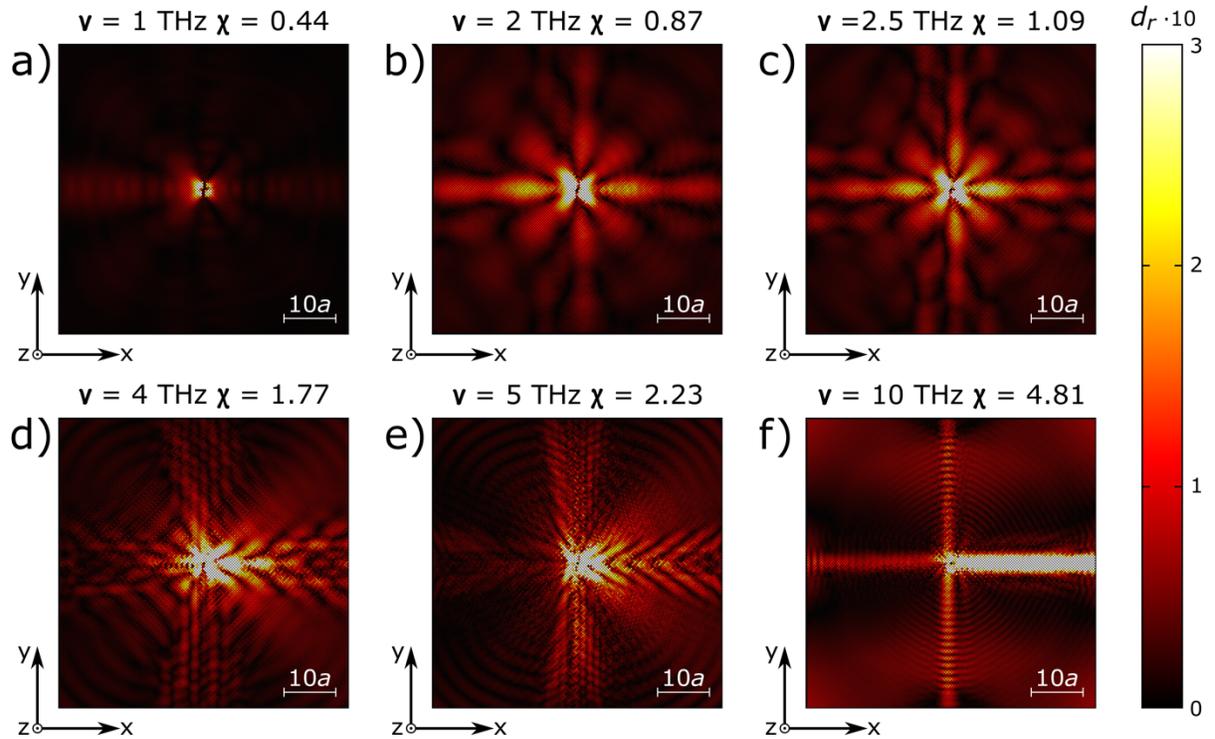

**Fig. S2.1.** The (x, y) slice of the spatial distribution of the radial component of the displacement scattered at a pore with a radius equal to 1$a$ for different modulation frequencies of the force: a) ν = 1 THz; b) ν = 2 THz; c) ν = 2.5 THz; d) ν = 4 THz; e) ν = 5 THz f) ν = 10 THz. k-vector of the incident wave is direct along the x-axis

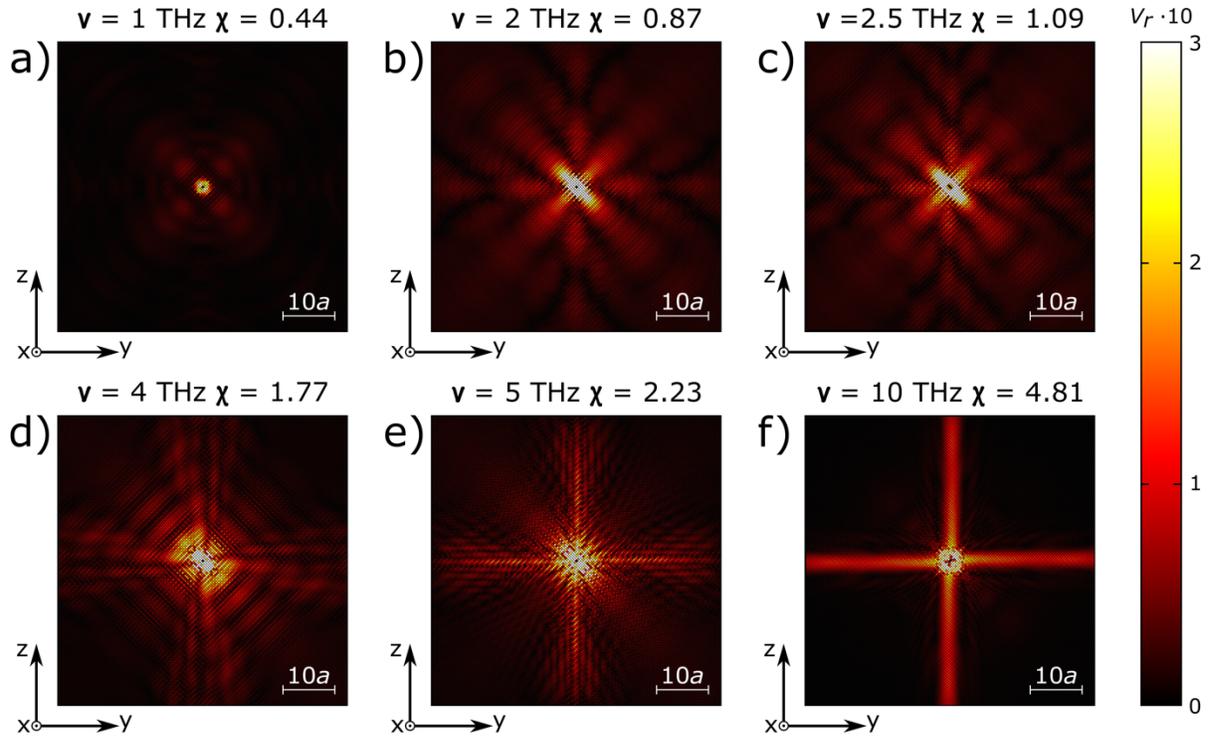

**Fig. S2.2.** The (y, z) slice of the spatial distribution of the radial component of the normalized velocity scattered at a pore with radius equal to 1*a* for different modulation frequencies of the force: a) ν = 1 THz; b) ν = 2 THz; c) ν = 2.5 THz; d) ν = 4 THz; e) ν = 5 THz f) ν = 10 THz. k-vector of the incident wave is direct along the x-axis

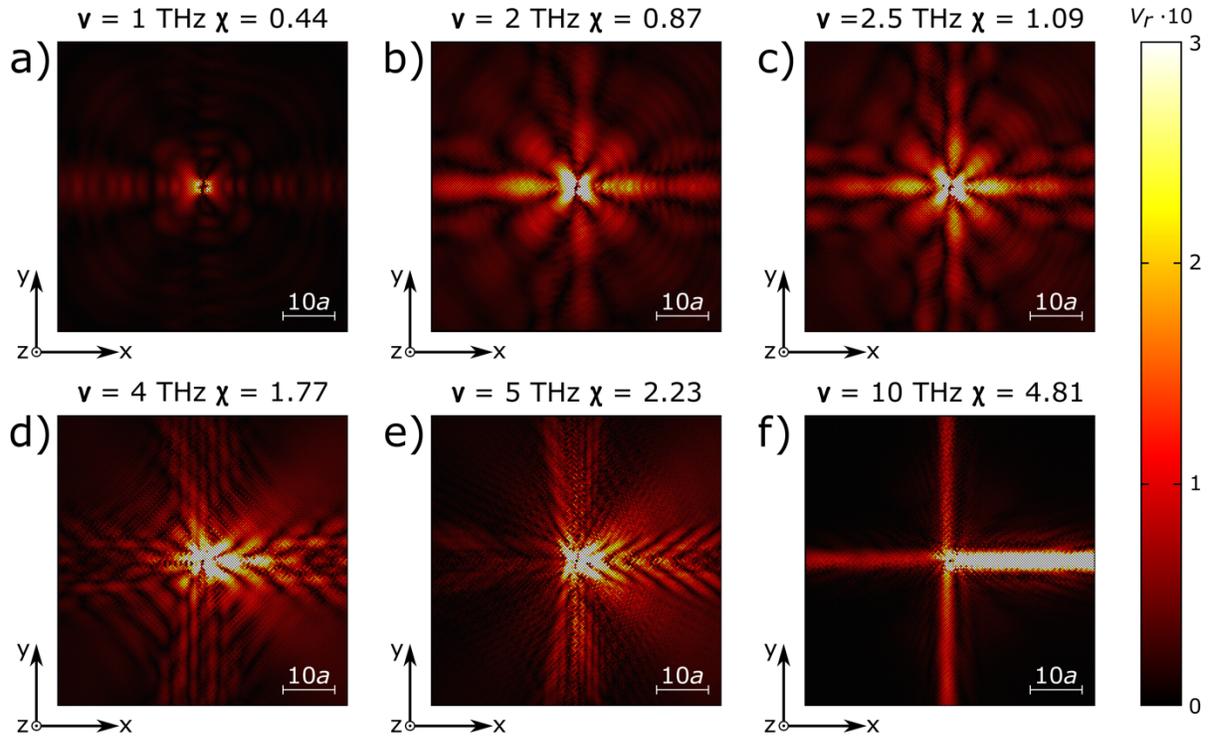

**Fig. S2.3.** The (x, y) slice of the spatial distribution of the radial component of the normalized velocity scattered at a pore with radius equal to 1$a$ for different modulation frequencies of the force: a) ν = 1 THz; b) ν = 2 THz; c) ν = 2.5 THz; d) ν = 4 THz; e) ν = 5 THz f) ν = 10 THz. k-vector of the incident wave is direct along the x-axis

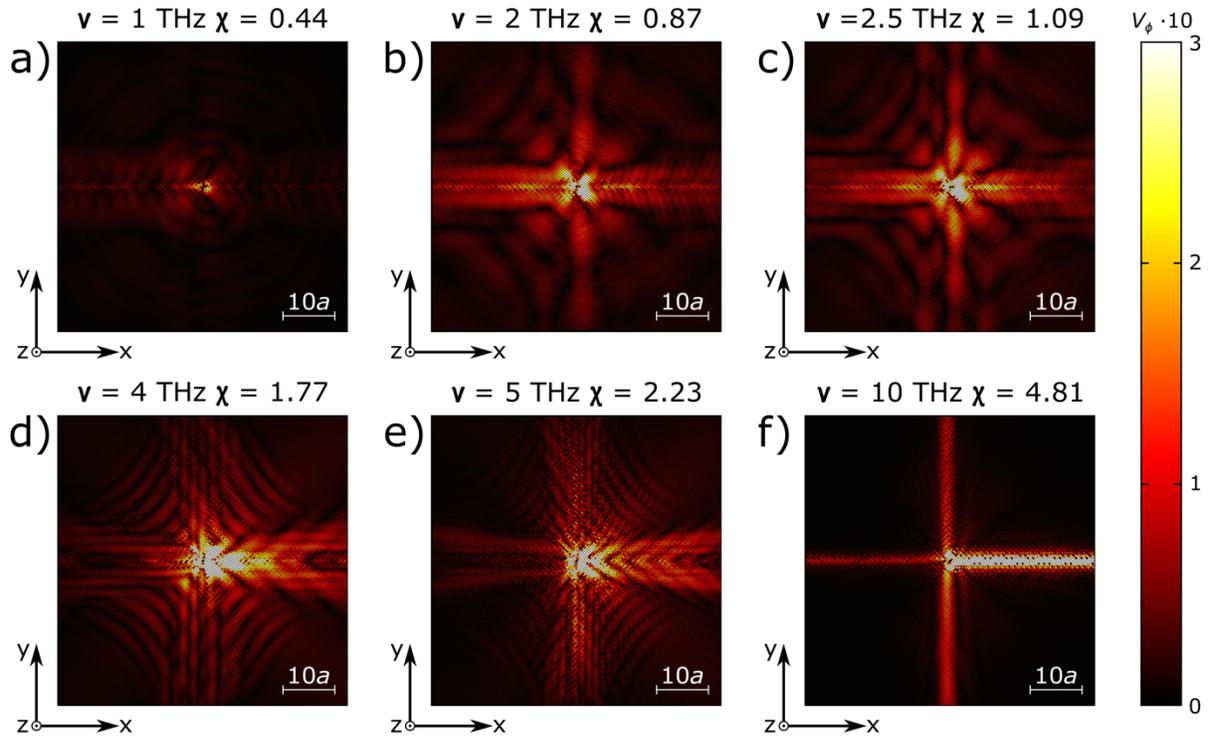

**Fig. S2.4.** The (x, y) slice of the spatial distribution of the polar component of the normalized velocity scattered at a pore with radius equal to 1$a$ for different modulation frequencies of the force: a) $v$ = 1 THz; b) $v$ = 2 THz; c) $v$ = 2.5 THz; d) $v$ = 4 THz; e) $v$ = 5 THz f) $v$ = 10 THz. k-vector of the incident wave is direct along the x-axis

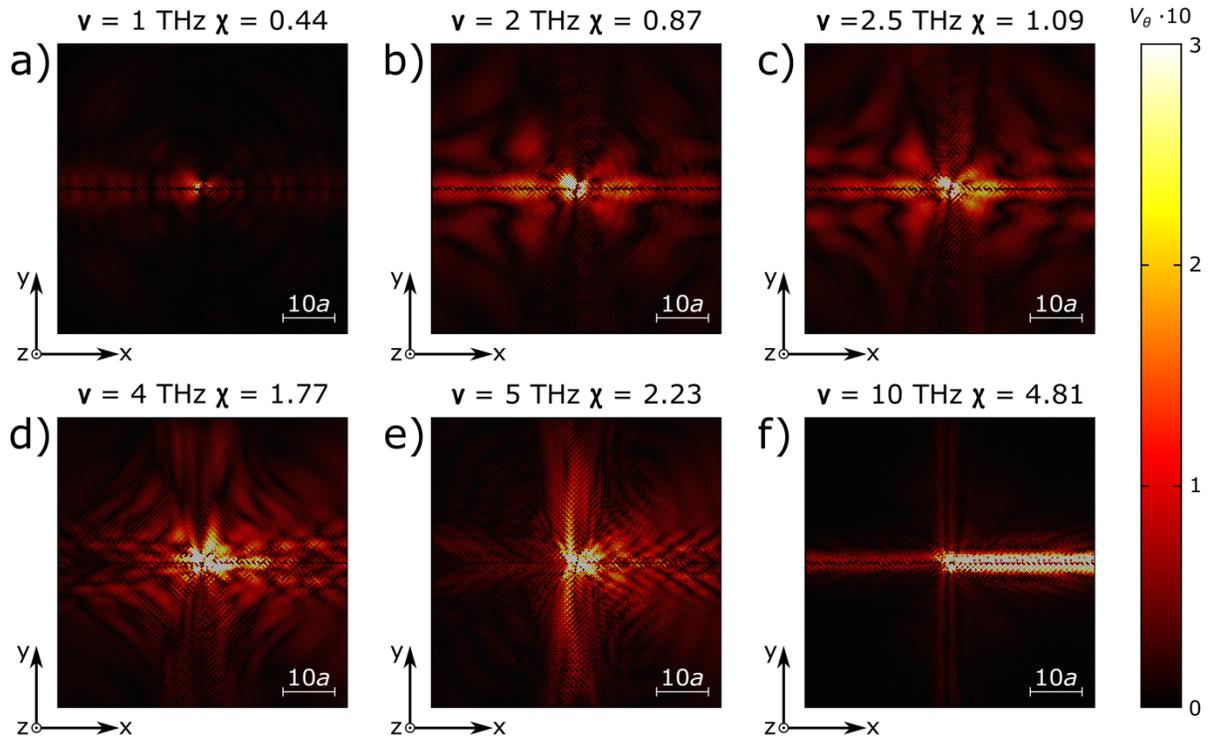

**Fig. S2.5.** The (x, y) slice of the spatial distribution of the azimuthal angle component of the normalized velocity scattered at a pore with radius equal to 1$a$ for different modulation frequencies of the force: a) v = 1 THz; b) v = 2 THz; c) v = 2.5 THz; d) v = 4 THz; e) v = 5 THz f) v = 10 THz. k-vector of the incident wave is direct along the x-axis

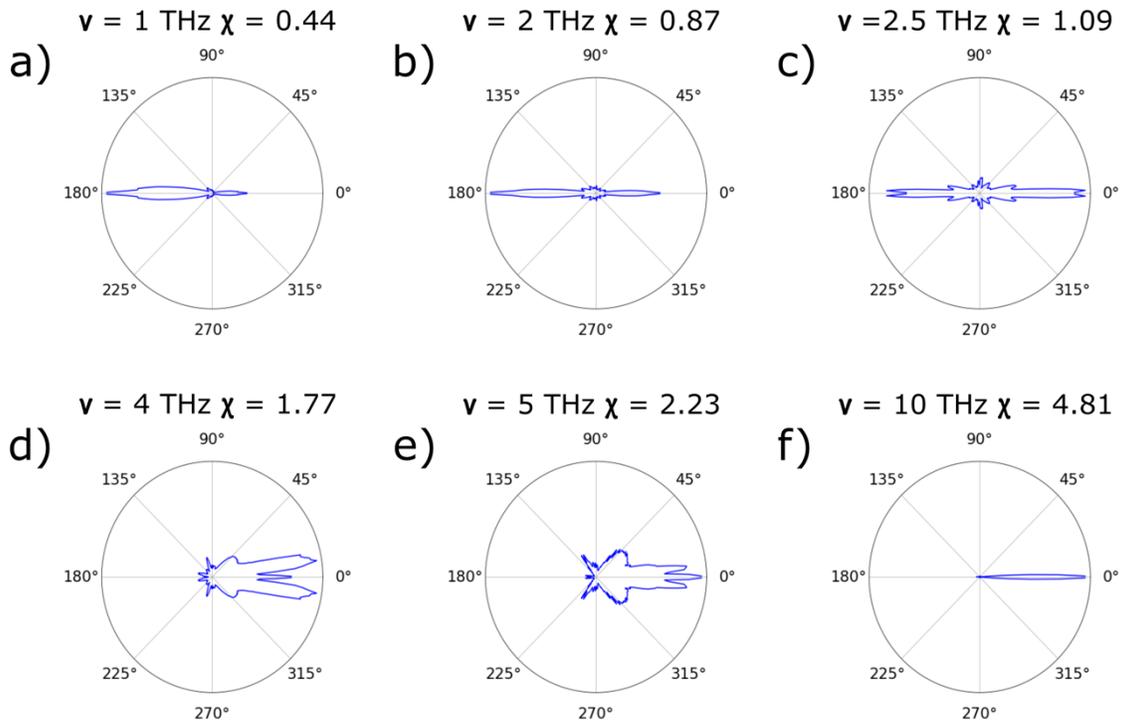

**Fig. S2.6.** The scattering diagram calculated with phase function ($R = 1a$) for the respective values of size parameter $\chi$: a) $\chi = 0.44$; b) $\chi = 0.87$; c) $\chi = 1.09$; d) $\chi = 1.77$; e) $\chi = 2.23$; f) $\chi = 4.81$.

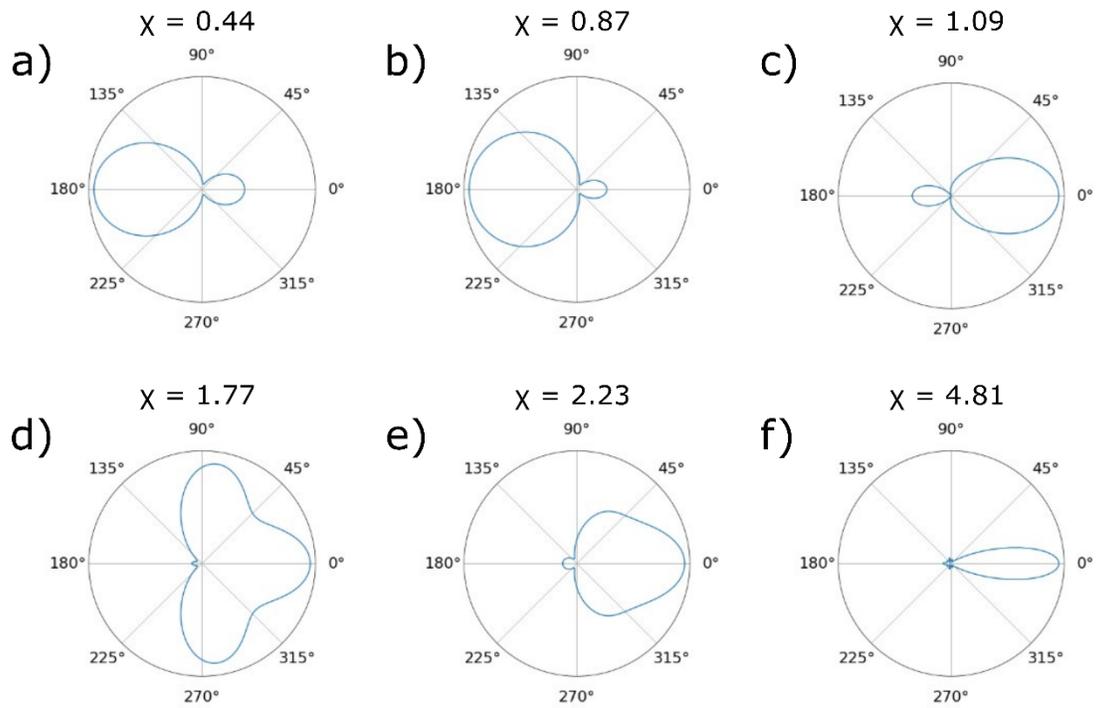

**Fig. S2.7.** The scattering diagram calculated with miepython for the respective values of size parameter χ: a) χ = 0.44; b) χ = 0.87; c) χ = 1.09; d) χ = 1.77; e) χ = 2.23; f) χ = 4.81. For definitenes we took complex index of refraction equal to m = 10 in the Mie model

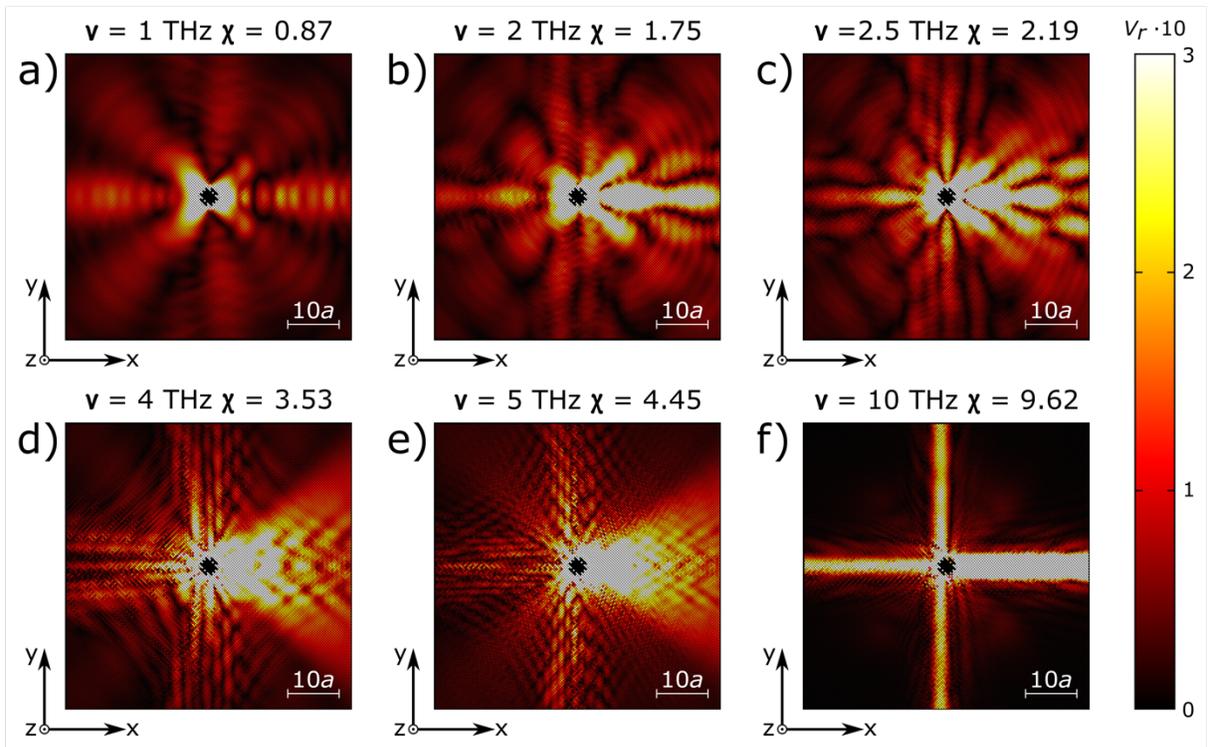

**Fig. S2.8.** The (x, y) slice of the spatial distribution of the radial component of the normalized velocity scattered at a pore with radius equal to 2*a* for different modulation frequencies of the force: a) ν = 1 THz; b) ν = 2 THz; c) ν = 2.5 THz; d) ν = 4 THz; e) ν = 5 THz f) ν = 10 THz. k-vector of the incident wave is direct along the x-axis

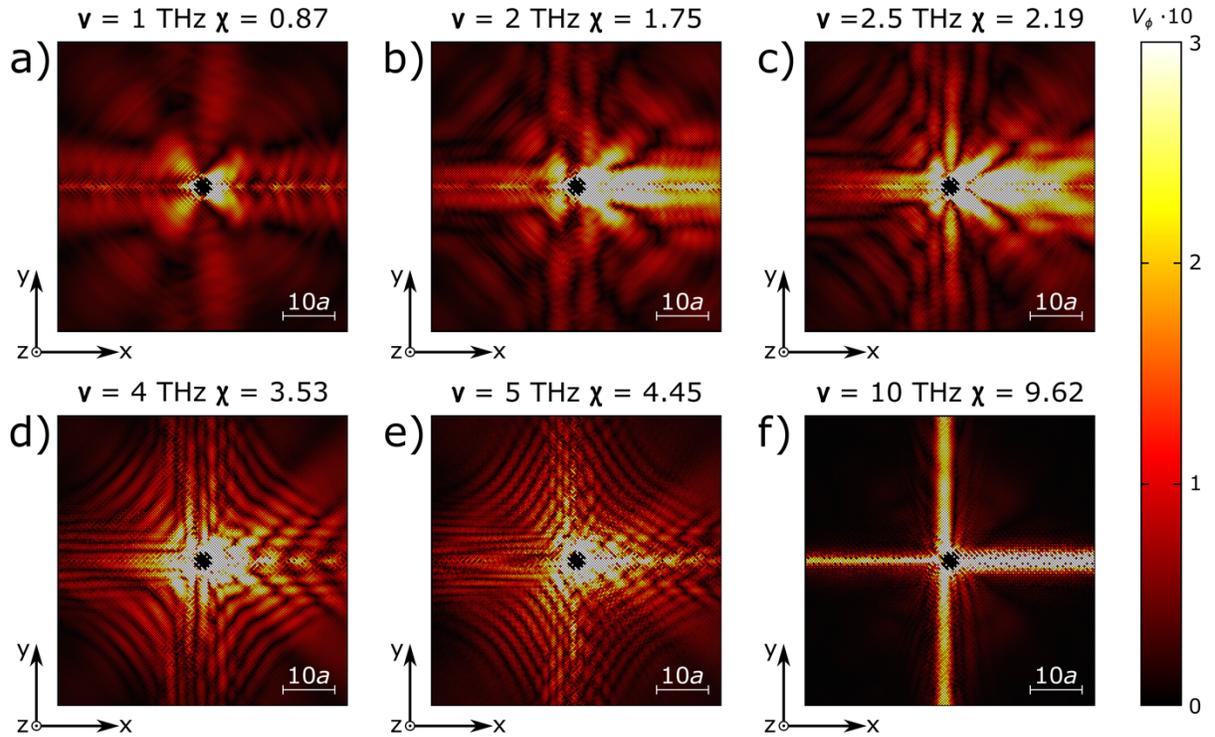

**Fig. S2.9.** The (x, y) slice of the spatial distribution of the polar component of the normalized velocity scattered at a pore with radius equal to $2a$ for different modulation frequencies of the force: a) $v = 1$ THz; b) $v = 2$ THz; c) $v = 2.5$ THz; d) $v = 4$ THz; e) $v = 5$ THz f) $v = 10$ THz. k-vector of the incident wave is direct along the x-axis

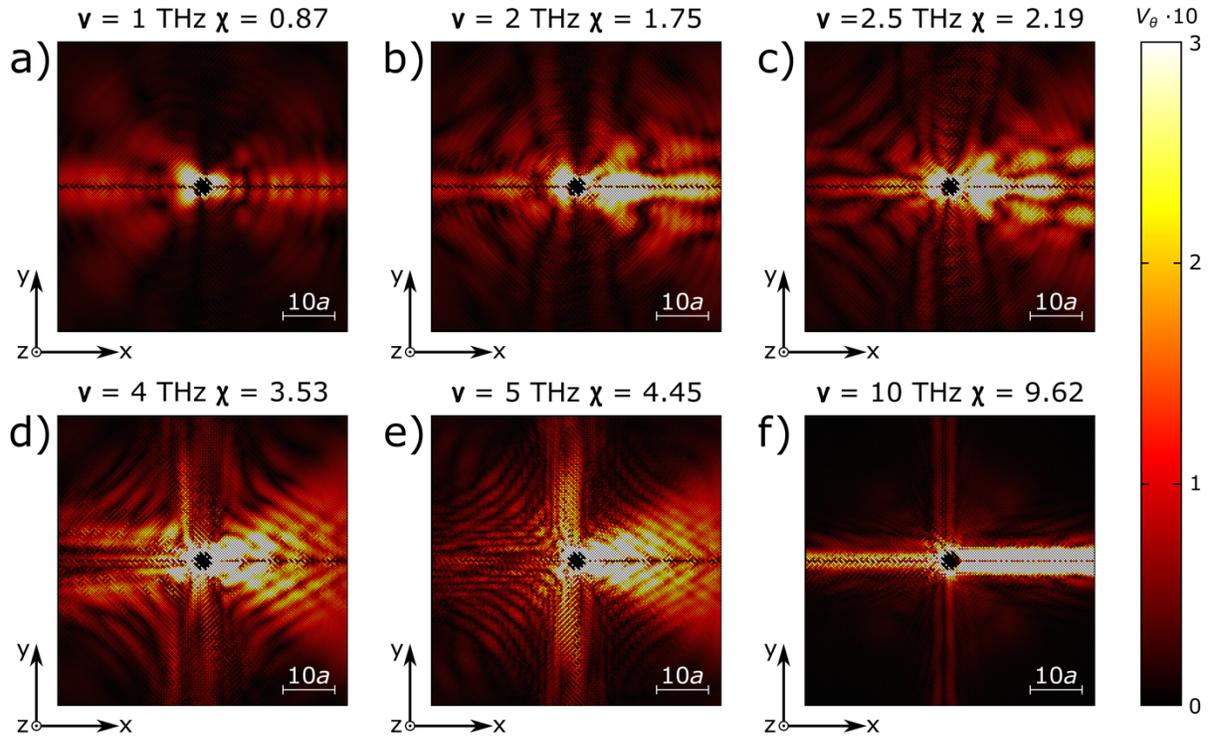

**Fig. S2.10.** The (x, y) slice of the spatial distribution of the azimuthal component of the normalized velocity scattered at a pore with radius equal to 2*a* for different modulation frequencies of the force: a) ν = 1 THz; b) ν = 2 THz; c) ν = 2.5 THz; d) ν = 4 THz; e) ν = 5 THz f) ν = 10 THz. k-vector of the incident wave is direct along the x-axis

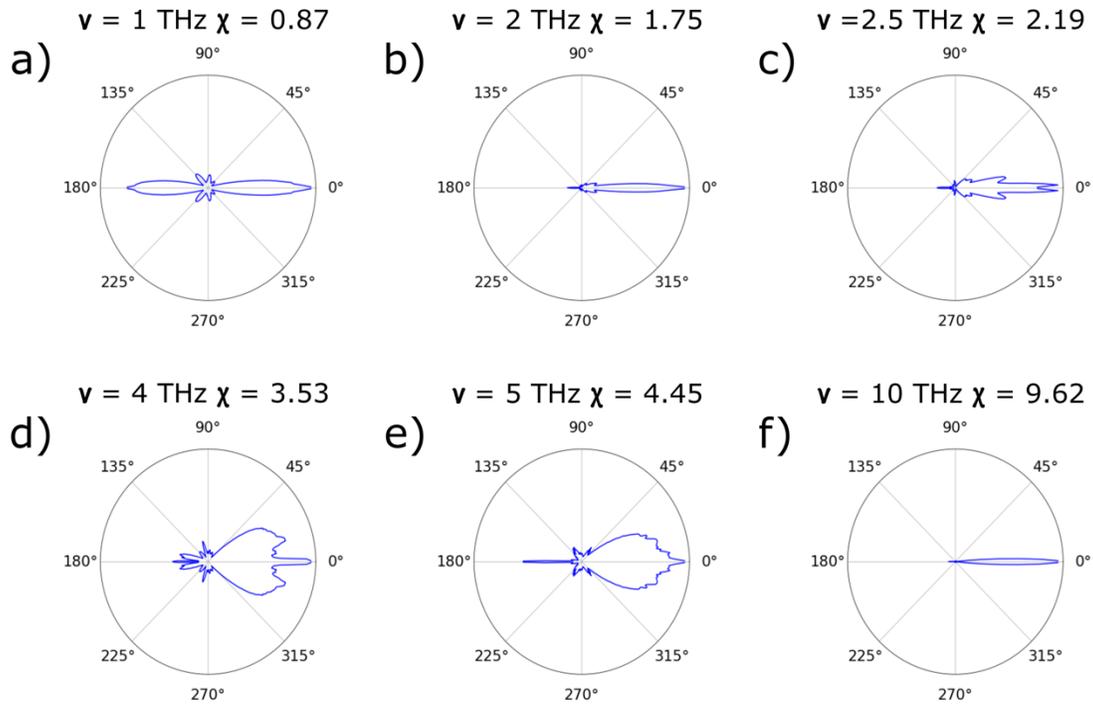

**Fig. S2.11.** The scattering diagram calculated with phase function ($R = 2a$) for the respective values of size parameter $\chi$: a) $\chi = 0.87$; b) $\chi = 1.75$; c) $\chi = 2.19$; d) $\chi = 3.53$; e) $\chi = 4.45$; f) $\chi = 9.62$.

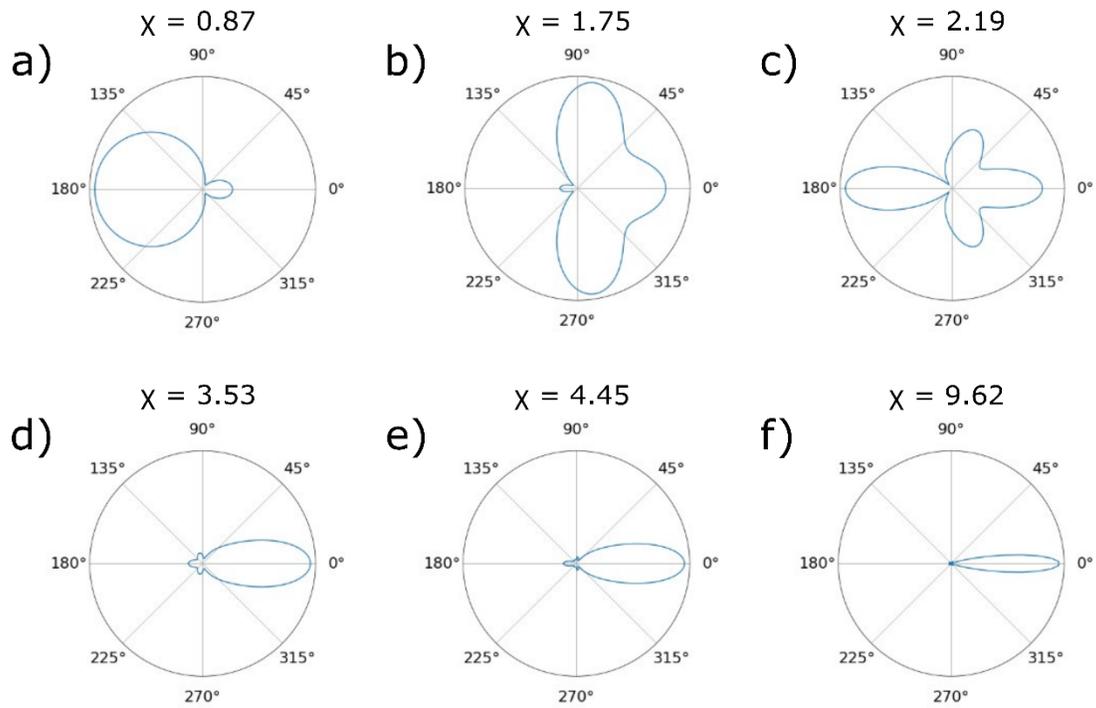

**Fig. S2.12.** The scattering diagram calculated with miepython for the respective values of size parameter χ: a) χ = 0.87; b) χ = 1.75; c) χ = 2.19; d) χ = 3.53; e) χ = 4.45; f) χ = 9.62. For definitenes we took complex index of refraction equal to m = 10 in the Mie model

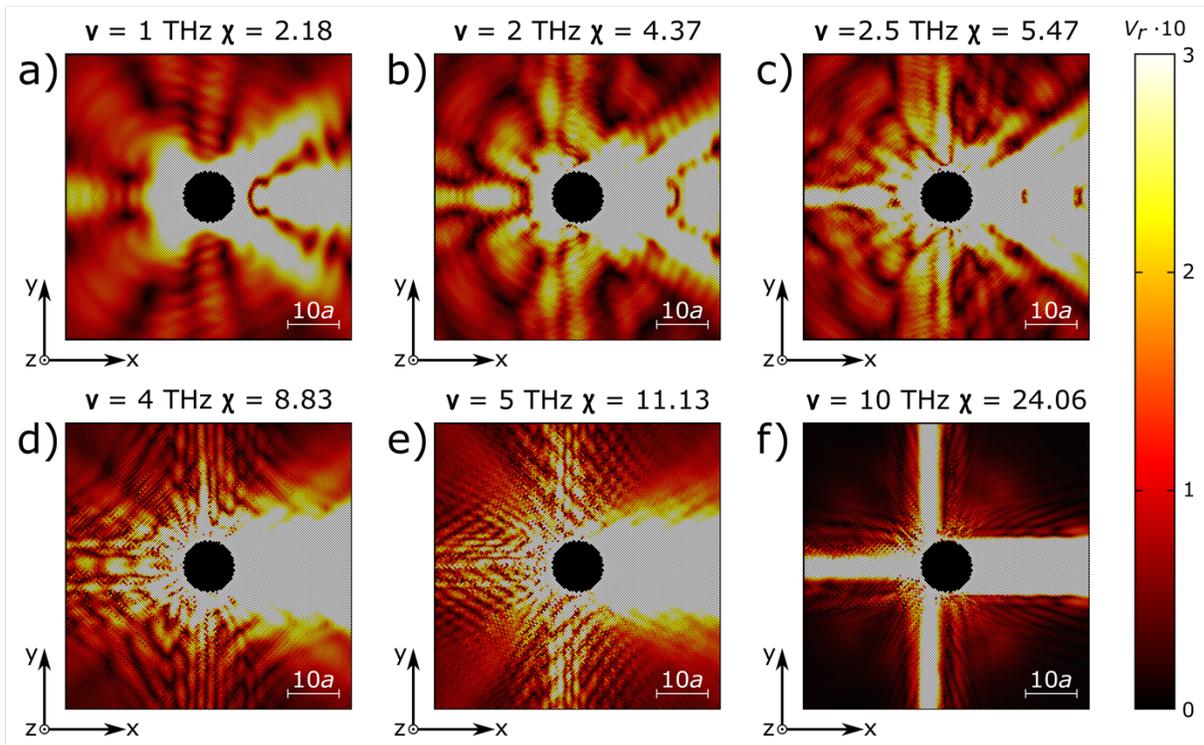

**Fig. S2.13.** The (x, y) slice of the spatial distribution of the radial component of the normalized velocity scattered at a pore with radius equal to 5*a* for different modulation frequencies of the force: a) ν = 1 THz; b) ν = 2 THz; c) ν = 2.5 THz; d) ν = 4 THz; e) ν = 5 THz f) ν = 10 THz. k-vector of the incident wave is direct along the x-axis

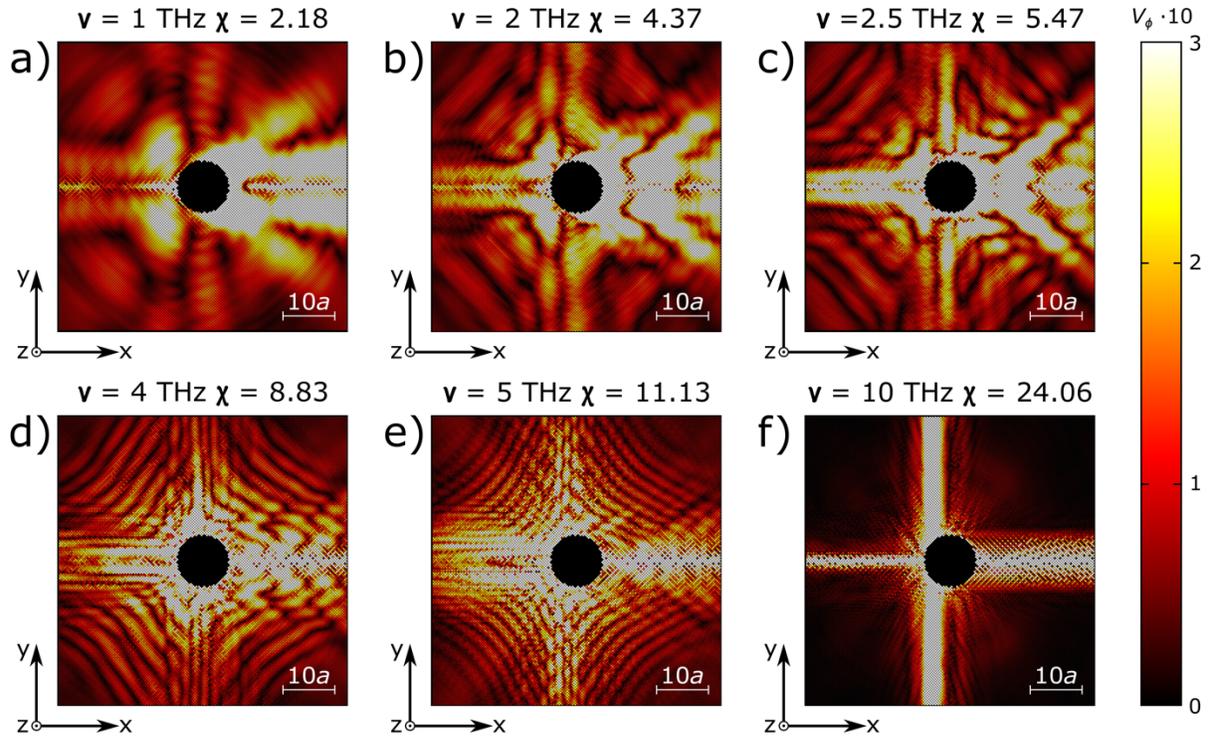

**Fig. S2.14.** The (x, y) slice of the spatial distribution of the polar component of the normalized velocity scattered at a pore with radius equal to 5$a$ for different modulation frequencies of the force: a) $v$ = 1 THz; b) $v$ = 2 THz; c) $v$ = 2.5 THz; d) $v$ = 4 THz; e) $v$ = 5 THz f) $v$ = 10 THz. k-vector of the incident wave is direct along the x-axis

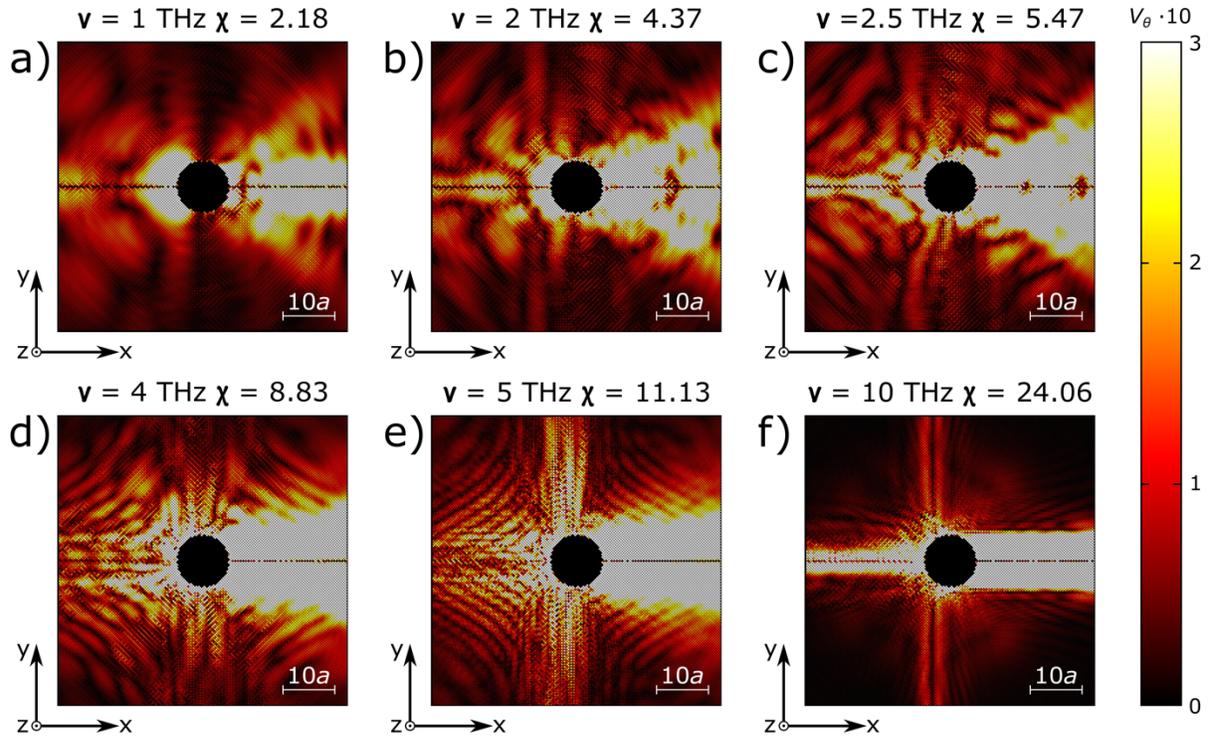

**Fig. S2.15.** The (x, y) slice of the spatial distribution of the azimuthal component of the normalized velocity scattered at a pore with radius equal to 5*a* for different modulation frequencies of the force: a) ν = 1 THz; b) ν = 2 THz; c) ν = 2.5 THz; d) ν = 4 THz; e) ν = 5 THz f) ν = 10 THz. k-vector of the incident wave is direct along the x-axis

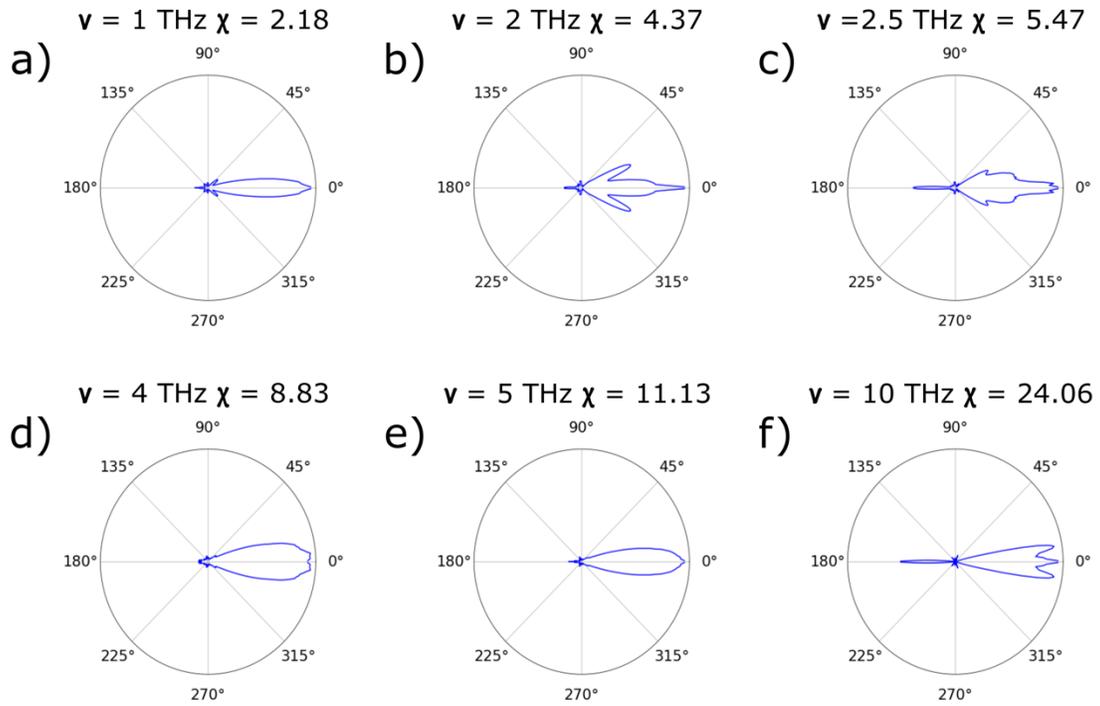

**Fig. S2.16.** The scattering diagram calculated with phase function ($R = 5a$) for the respective values of size parameter $\chi$: a) $\chi = 2.18$; b) $\chi = 4.37$; c) $\chi = 5.47$; d) $\chi = 8.83$; e) $\chi = 11.13$; f) $\chi = 24.06$.

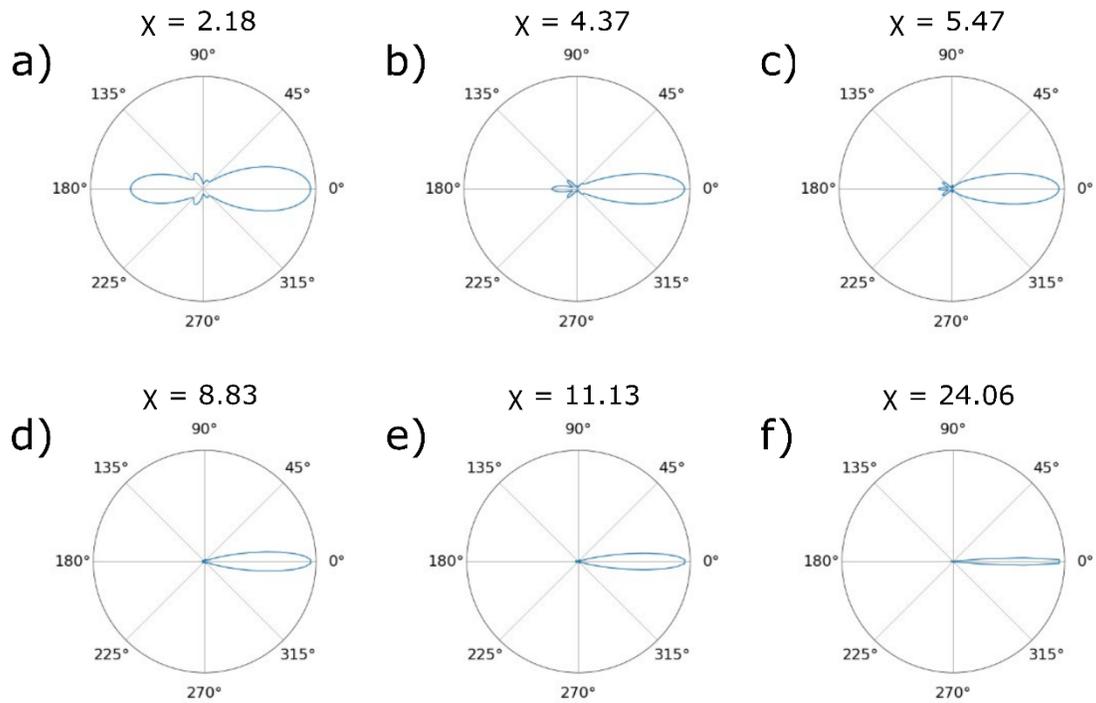

**Fig. S2.17.** The scattering diagram calculated with miepython for the respective values of size parameter χ: a) χ = 2.18; b) χ = 4.37; c) χ = 5.47; d) χ = 8.83; e) χ = 11.13; f) χ = 24.06. For definitenes we took complex index of refraction equal to m = 10 in the Mie model

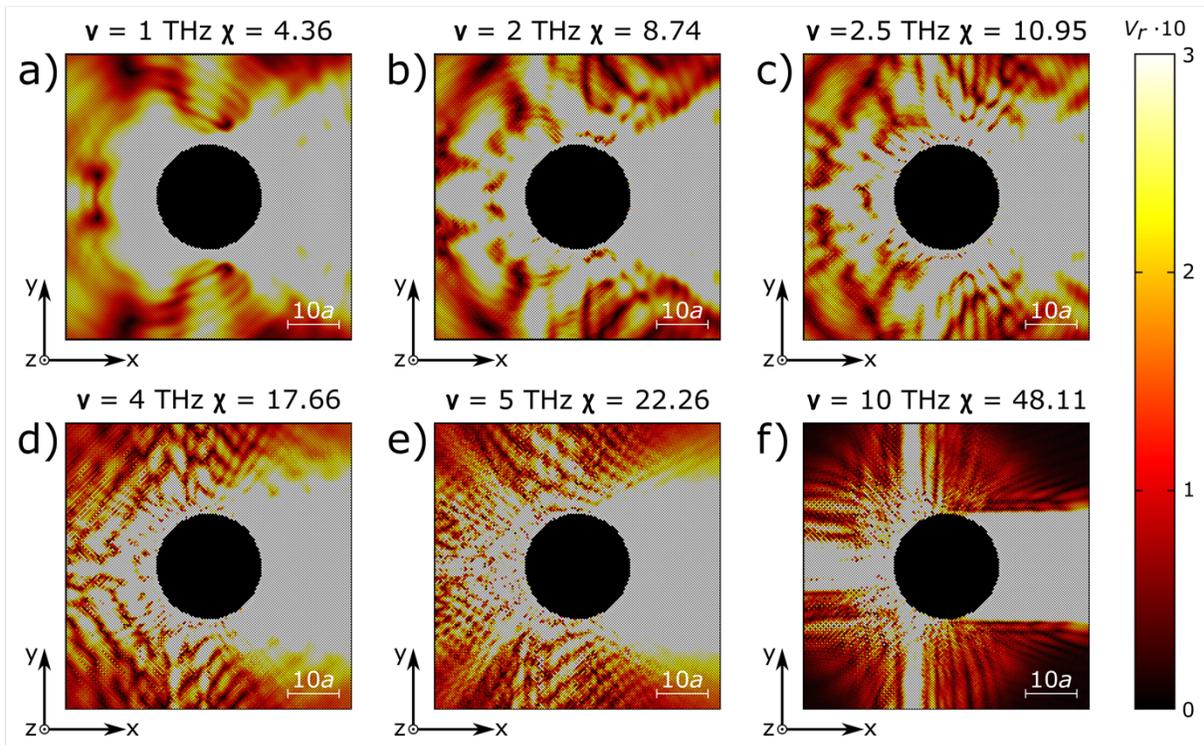

**Fig. S2.18.** The (x, y) slice of the spatial distribution of the radial component of the normalized velocity scattered at a pore with radius equal to 10$a$ for different modulation frequencies of the force: a) ν = 1 THz; b) ν = 2 THz; c) ν = 2.5 THz; d) ν = 4 THz; e) ν = 5 THz f) ν = 10 THz. k-vector of the incident wave is direct along the x-axis

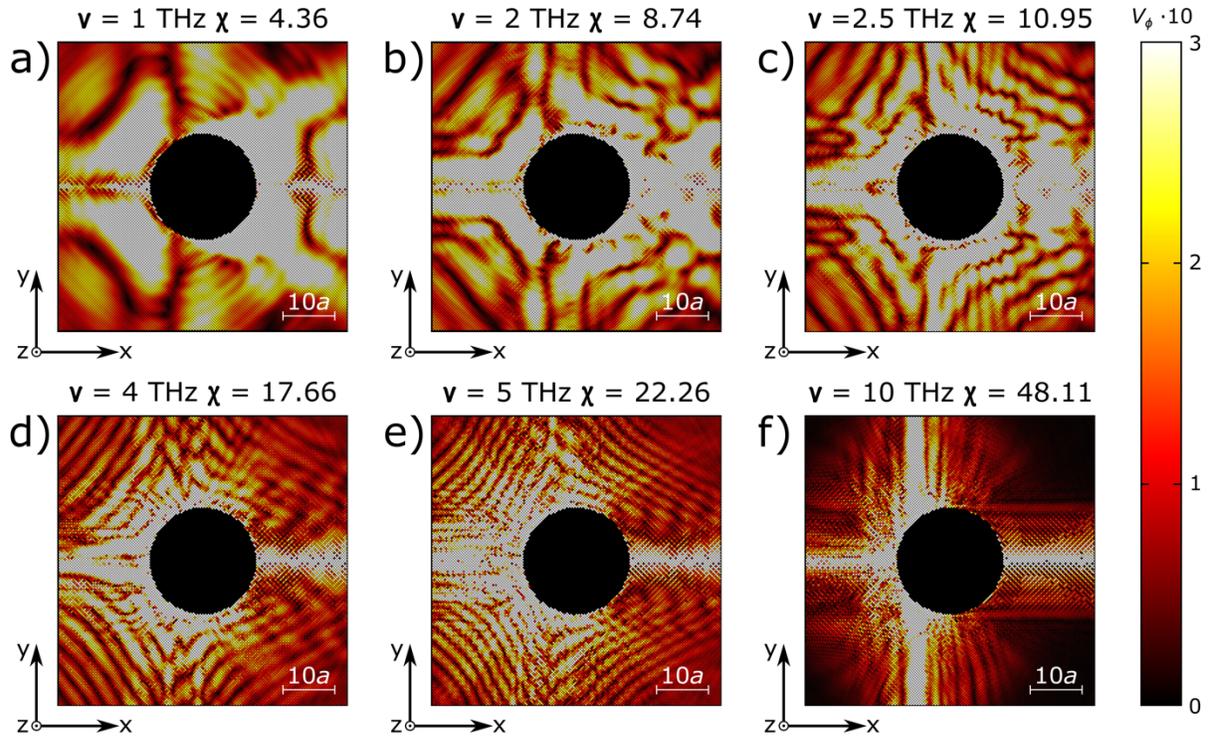

**Fig. S2.19.** The (x, y) slice of the spatial distribution of the polar component of the normalized velocity scattered at a pore with radius equal to 10*a* for different modulation frequencies of the force: a) ν = 1 THz; b) ν = 2 THz; c) ν = 2.5 THz; d) ν = 4 THz; e) ν = 5 THz f) ν = 10 THz. k-vector of the incident wave is direct along the x-axis

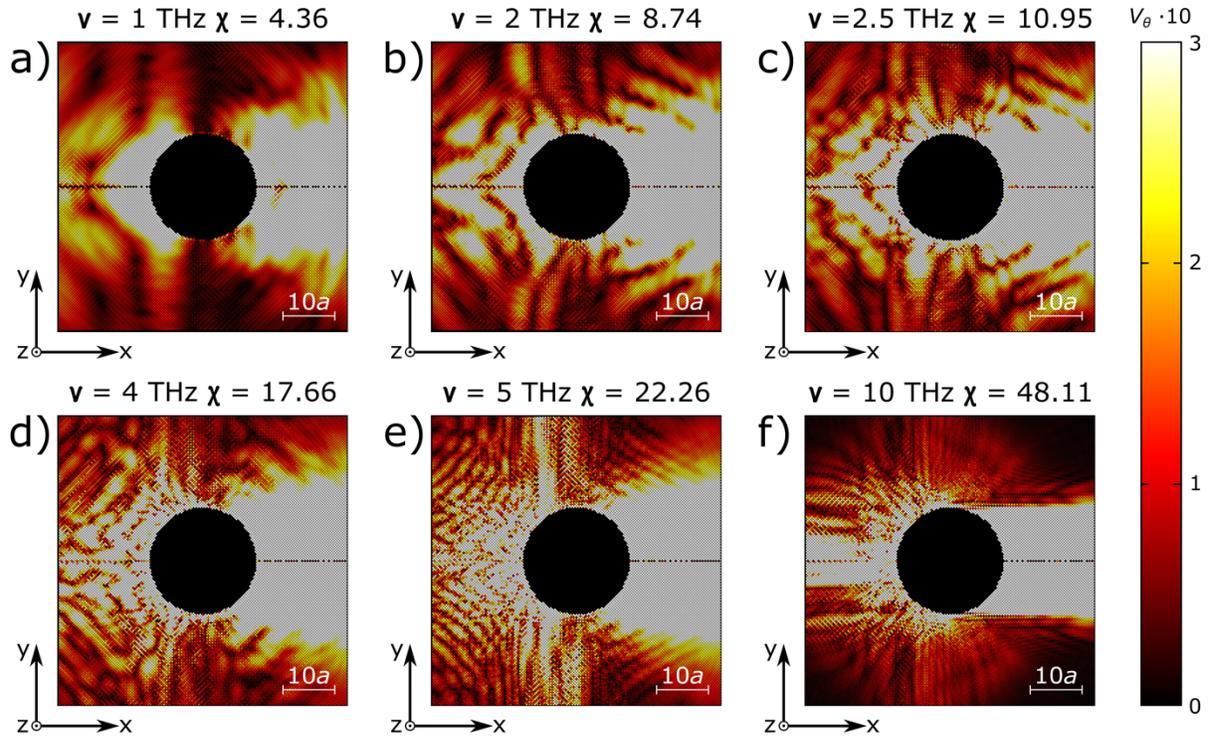

**Fig. S2.20.** The (x, y) slice of the spatial distribution of the azimuthal component of the normalized velocity scattered at a pore with radius equal to 10$a$ for different modulation frequencies of the force: a) ν = 1 THz; b) ν = 2 THz; c) ν = 2.5 THz; d) ν = 4 THz; e) ν = 5 THz f) ν = 10 THz. k-vector of the incident wave is direct along the x-axis

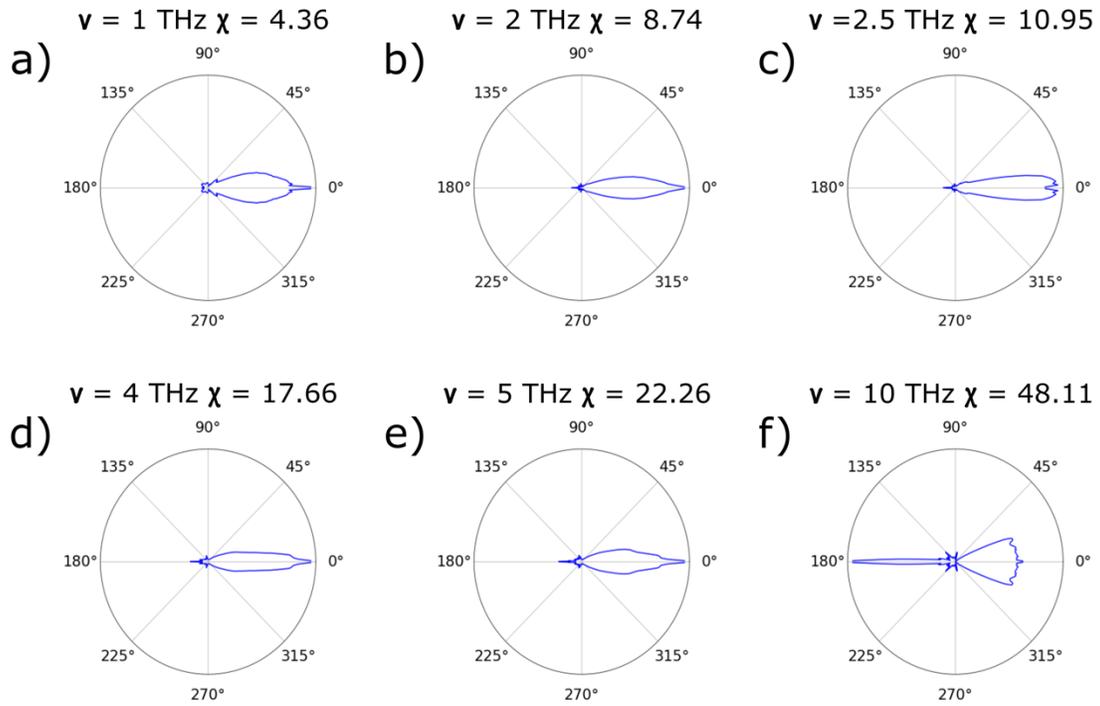

**Fig. S2.16.** The scattering diagram calculated with phase function ($R = 10a$) for the respective values of size parameter $\chi$: a) $\chi = 4.36$; b) $\chi = 8.74$; c) $\chi = 10.95$; d) $\chi = 17.66$; e) $\chi = 22.26$; f) $\chi = 48.11$. For f) case we observed the wide angular forward scattered galo, which arises due to the finit radius of data avareging (in Mie theory it should be much bigger than the pore radius). Since the calulation difficalties we could not calculate ath higher distance, but we observed the clear trend of its narrowing with distance

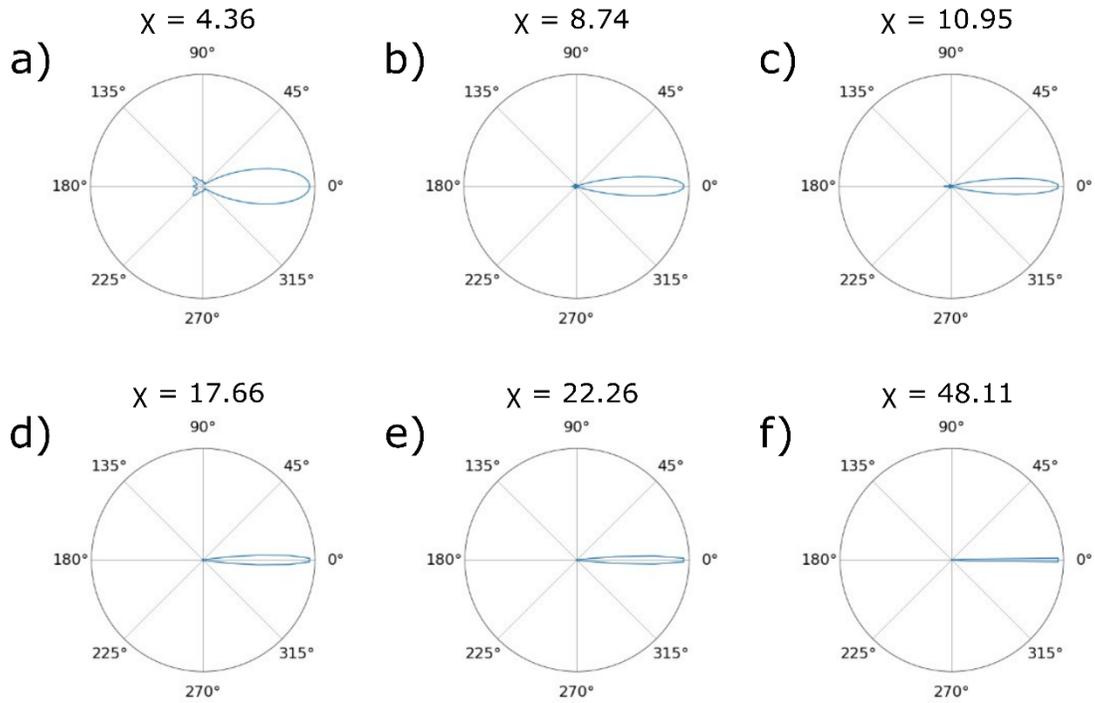

**Fig. S2.17.** The scattering diagram calculated with miepython for the respective values of size parameter χ: a) χ = 4.36; b) χ = 8.74; c) χ = 10.95; d) χ = 17.66; e) χ = 22.26; f) χ = 48.11. For definitenes we took complex index of refraction equal to m = 10 in the Mie model